\documentclass[12pt]{article}
\title{Comment on ``Scientific Regress'' ({\it First Things\/} May 2016)}
\author{J. I. Katz \\ Dept. Physics, Washington University, St. Louis, Mo. 63130}
\date{\today}
\begin{document}
\maketitle
\begin{abstract}
Response to William A. Wilson on the limits and fallibility of science.
\end{abstract}
                    
     William A. Wilson points to many of the ills of contemporary science.
Some of them are ineluctable features of human nature; science is done by
scientists, who have all the flaws of human beings.  In the search of
reputation and career they become ``success-oriented'', weighting evidence
in favor of their discoveries more than that against.  The same psychology
that leads a soldier to suppress fear leads the scientist to suppress
skepticism, but what is necessary for the soldier is destructive in a
scientist.

     Human nature cannot be changed, but the conditions under which it 
operates can.  Excessive competitive pressures lead to corner-cutting.
That has simple solutions: train fewer scientists, spread research support
more broadly and thinly, and abandon the pretense that the quality of
scientific research can be quantified in the same manner as the phenomena
the scientist observes.  Award the same kudos to the scientist who disproves
a previously accepted result as to one who made the original, now
discredited, discovery.

     But there is more that needs to be done.  Almost every paper in the
branches of science with the most irreproducible results contains a formal
statistical analysis ``demonstrating'' its significance.  Biomedical
research teams generally include a professional biostatistician.  No paper
can be published without a formal demonstration of its statistical validity,
and the chance of its result being the product of random chance must be less
than 5\%, and often turns out to be very much less.

     Yet roughly half of such results are not reproducible, and often the
problem is not that the original researchers didn't allow for the large
number of possible hypotheses (stones in the field, in Wilson's article).
Biostatisticians are neither stupid nor dishonest.  The problem is that
formal statistical analysis cannot allow for systematic bias.  Perhaps the
apparent benefits of a daily glass of wine result not from the wine, but 
from something else: perhaps social wine drinkers are the kind of people who
follow doctors' orders, for example.  In physics there is a saying that half
of all ``three sigma'' results are wrong; that would happen 0.3\% of the
time if statistical analysis described the real world.

     In more theoretical branches of science a different problem arises, the
use of elaborate computer codes to model complex phenomena.  Modern codes
contain our best understanding of many interacting processes, derived from
fundamental principles or simpler experiments.  Yet these interact in poorly
understood ways, making inferences uncertain.  The National Ignition
Campaign's calculations, based on empirical data and fundamental physical
laws, predicted that a pellet of fusion fuel would ignite, but it didn't.

     The problem is worse in fields like climate that lack the vast base of
data available to fusion researchers; climate models are calibrated by our
single climate.  The physics of greenhouse gases and warming was understood 
qualitatively in the 1890's, but the predictive power of even the best
modern calculations remains uncertain; different groups make predictions
that disagree by a factor of two.  This is perfect ground for the Cult of
Science, its leaders lured by the scent of possible catastrophe (who
wouldn't like to save humanity?), advertising the latest untestable
predictions and demanding action NOW.

Science gets some very important things right.  We can send a probe to Mars
and have it arrive within a few hundred feet of target.  We can analyze the
DNA of a microbe, or of people, and determine the history of their 
ancestors.  And much more.  But unless we recognize the limits of our
methods and our understanding, we will be led astray.
\end{document}